*Article*

# Preprocessing for Lessening the Influence of Eye Artifacts in EEG Analysis

**Alejandro Villena, Lorenzo J. Tardón *****, Isabel Barbancho *****, Ana M. Barbancho**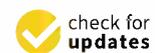**, Elvira Brattico † and Niels T. Haumann †**

ATIC Research Group, E.T.S.I. Telecomunicación, Universidad de Málaga, Andalucía Tech, Campus Universitario de Teatinos s/n, 29071 Málaga, Spain; avillenarod@gmail.com (A.V.); abp@ic.uma.es (A.M.B.); elvira.brattico@gmail.com (E.B.); niels.haumann@clin.au.dk (N.T.H.)
* Correspondence: lorenzo@ic.uma.es (L.J.T.); ibp@ic.uma.es (I.B.);
  Tel.: +34-952131188 (L.J.T.); +34-952132587 (I.B.)
† Current address: Department of Clinical Medicine, Center for Music in the Brain, Nørrebrogade 44, Building 10G, 8000 Aarhus C, Denmark.

**Abstract:** We dealt with the problem of artifacts in EEG signals in relation to the usage of lengthy trials. Specifically, we considered eye artifacts found in EEG signals, their influence in the analysis of the data and alternatives to diminish their impact on later studies of brain activity on lengthy tasks. We proposed a scheme of partial rejection of independent signal components, provided a method to extract EEG signal components with diminished influence of eye artifacts, and assess the importance of using artifact free signal excerpts to extract signal components in order to analyze brain activity in a musical context.

## 1. Introduction

Artifacts like saccades, eye blinks, muscle noise, heart signals, or even line noise are present in most electroencephalography (EEG) recordings. They produce voltage fluctuations that are registered by the electrodes placed in the scalp and contaminate the brain signals measured by EEG. Consequently, the identification, cancellation, or correction of such artifacts is often required. Most of the artifacts coming from the contraction of muscles contain frequencies above 100 Hz and because of this, electromyographic (EMG) activity can be eliminated by suppressing frequencies above 100 Hz [1]. Power line interference can be removed by making use of notch filters [2]. For other complex artifacts like eye blinks or eye movements, artifact rejection or correction methods are required to diminish their influence on the EEG signals [3].

Trial-wise rejection based on the presence of artifacts is a crude but valid process for short trial experiments in which a select subset with the contaminated trials is eliminated [1]. However, a healthy person can blink up to 20 times per minute in resting conditions [4] which means that such an artifact rejection strategy cannot be used for long trial experiments. This is the case of trials that imply listening to real world music excerpts; it turns out to be impossible for a subject not to blink or move their eyes at all during a trial.

When it comes to blink and saccade correction, several methods have been suggested such as time domain regression [5], frequency domain regression [6], Kalman filtering [7], or singular value decomposition (SVD) [8]. However, because EEG and ocular activity mix bidirectionally, using regression for eye artifacts inevitably produces the loss of a relevant EEG signal [9]. More recently, unmixing algorithms like principal component analysis (PCA) [10] or independent component analysis (ICA) [9,11–13] have been used to separate the underlying signals present in the multichannel EEG recordings. Generally, these studies show that PCA cannot completely separate eye artifacts from brain signals due to the PCA assumption according to which the decomposed components are algebraically orthogonal, which is in general difficult to satisfy [14]. ICA methods, on their side, are based on the assumption that the observed signals are a linear combination of $n$ unknown and statistically independent sources [13]. The goal of ICA is to find a separating or unmixing matrix that extracts the independent components from the signals. With regard to EEG recordings, this model implies that the signals recorded on the scalp are mixtures of time courses of temporally independent cerebral and artifactual signals in such a way that potentials arising from different parts of the brain and diverse activities are summed linearly at the electrodes with propagation delays negligible [9].

The main use of the ICA approach to the eye artifact removal problem relies on the elimination of the artifactual components before mixing the components back to obtain clean EEG multichannel data.

However, which components or excerpts should be considered artifactual and removed and which ones should not, must be defined. This question is usually answered by a visual inspection of the waveform and scalp topographies and the manual selection by an experienced scientist [9,11]. However, Li et al. [13] proposed an automatic method for the selection of artifact-related components by matching the given scalp distribution of each component to a previously generated template; Joyce et al. describe [14] an automatic process for eye movement removal relying on the elimination of components based on the correlation with the electrooculogram (EOG) channel. Machine-learning approaches have also been proposed for the identification of artifacts in EEG signals after ICA [15] and the utilization of convolutional neural networks (CNNs) [16] has lately become possible with advances in computation.

In this manuscript, we present and compare two approaches for the treatment of artifactual independent components (ICs) generated by ICA. In both methods, the identification and selection of the artifact-related ICs is based on the cross-correlation coefficient between the channel and each ICs. The most correlated-to-EOG ICs are selected as odd ones. In the first approach, the samples affected by the artifacts in the selected ICs are attenuated before mixing the components back into the EEG data channels. In the second method, we aim to obtain an unmixing matrix for ICA with the influence of eye artifacts diminished; to this end, the samples in all the EEG channels in the intervals when artifacts are encountered are removed before the ICA algorithm. Such an approach will produce a different unmixing matrix in which the influence of eye artifacts is reduced.

This paper is organized as follows: Section 2 describes the experimental setup details such as the participants data, the stimuli used, and the equipment and procedure followed to acquire the EEG recordings. Section 3 introduces the data preprocessing scheme. In Section 4, the novel partial rejection method and the artifact-diminished unmixing matrix method are described in detail. Section 5 presents the experimental results and a comparison of the ICs complete removal method versus the ICs partial rejection method, together with a discussion of the significance and limitations of the proposed methods, including the influence of artifacts on the unmixing matrix. Finally, Section 6 presents the conclusions extracted from this work.

## 2. Experimental Setup

In this section, the methodology followed to record EEG signals during music listening is described. This includes the description of the subjects participating in the experiments, the stimuli, and the specific equipment and recording procedure employed.

### 2.1. Participants

A total of 5 right-handed native Spanish speakers took part in the experiment: 3 female and 2 male subjects, aged 20 to 29 years (22.8 on average) volunteered. Participants claimed that they did not suffer from hearing loss nor had a history of neurological illnesses. They also, confirmed not to be under psychopharmacology treatments or pregnant, and had no metal pieces allocated in their skull.

Additionally, all participants were asked to confirm the non-use of dreadlocks or a similar haircut and lacquer or any kind of hair gel. Table 1 shows the age, sex, and number of final clean channels measured for every participant. Bad channels, in which abnormally high or low signal amplitudes were visually identified, were removed.

Every subject gave their written consent for the tests according to the ethical committee of the Universidad de Málaga with PEIBA code 1403-N-17.

**Table 1.** Participants' data.

| Id | Age | Sex | Number of Clean Channels |
|----|-----|-----|--------------------------|
| $S_9$ | 21 | Female | 12 |
| $S_{10}$ | 21 | Female | 12 |
| $S_{12}$ | 29 | Male | 14 |
| $S_{18}$ | 23 | Male | 14 |
| $S_{19}$ | 20 | Female | 11 |

### 2.2. Stimuli

Four long pieces of instrumental music from different genres were presented to the participants as stimuli: A modern tango (Adios Nonino by Astor Piazzolla), an opera (The Magic Flute by Mozart), a waltz (Emperor Waltz by Johann Strauss), and a progressive rock track (Universal Mind by Liquid Tension Experiment) as well as a final piece of 1 min 20 s of speech in their native language.

The 8 min 30 s tango of Astor Piazzolla was recorded in a concert in Lausanne Switzerland; due to the live conditions of the track, the first few seconds were cropped to avoid the applauses from the audience, making it a final length of 8 min 7 s. This version of the track is the same as the one used by Poikonen et al. in [17]. The other 3 music tracks were employed as originally recorded; they remained untouched. The speech track contains 6 unrelated readings from Spanish native speakers without any other music or sound. Track durations and their relation to the experiment's trial numbers are shown in Table 2.

**Table 2.** Track lengths and their correspondence with the trial numbers.

| Track | Trial | Duration (in min:s) |
|-------|-------|---------------------|
| The Magic Flute | 1 | 7:13 |
| Emperor Waltz | 2 | 5:31 |
| Universal Mind | 3 | 7:54 |
| Adios Nonino | 4 | 8:07 |
| Speech | 5 | 1:20 |

### 2.3. Equipment and Procedure

The procedure employed follows the directions described by Poikonen et al. in [17]. The stimuli were presented to the participants by using the software piece Presentation via Sennheiser HD219 headphones. An individual hearing threshold was determined for each participant by playing the royalty free track 'Casual Friday' by Bjokib https://bjokib.com/casual-friday-royalty-free-music-free-download/, which was irrelevant to the study. The procedure employed to determine such thresholds were: The playback loudness level started clearly above the hearing threshold and it was manually decreased until the participants reported they stopped hearing the sound; then, the test track was played clearly below the hearing threshold and the loudness was manually increased until the participants reported hearing the sound. The average between these two loudness levels was considered the hearing threshold. After that, the playback loudness level was set to 50 dB above the individually determined hearing threshold.

The audio tracks were played in the following order: Magic Flute, Emperor Waltz, Universal Mind, Adios Nonino, and, lastly, the speech track. The resting period between the pieces of music was determined by the subjects themselves. Each track started after pressing a button whenever they were still, comfortable, and ready to listen to the next piece. Once the button was pressed, the audio track started after an informative message and 3 s of silence.

The EEG data were recorded by using OpenVibe v2.2.0 software with a 16-channel TMSi Porti system with water-based electrodes. A total of 14 electrodes were placed on the scalp according to the modified combinatorial nomenclature of the 10–20 system for the placement of electrodes [18]; the specific electrode locations employed were: Fz, F3, F4, FCz, Cz, C3, C4, T7, T8, CPz, Pz, P3, P4, POz. Furthermore, one electrode was placed vertically at the right eye (EOG) and one as ground on the forehead. The TMSi system takes the average of all electrodes as reference. The data were collected at a 2048 Hz sampling rate. The beginning and end of each musical piece was marked with a trigger into the EEG data.

The EEG data of all the participants were imported into Matlab 2018b [19] and the processing was made mostly with the FieldTrip toolbox [20].

## 3. Signal Preprocessing

In this section, the specific processing stages considered in this work, regarding trial detection and filtering, are described; recall that bad channels were previously removed by visually inspecting the EEG data. Also, some comments on the process of annotation of eye artifacts are given.

### 3.1. Trial Detection

The EEG multichannel data $\mathbf{x}(t) = [x_1(t), x_2(t), \ldots, x_n(t)]^T$ are recorded as a continuous stream that is sliced up into the trials of the experiment according to Table 2.

Accurate partition is possible thanks to the auxiliary trigger channel that contains several narrow steps that mark the beginning and end of every song in the experiment, plus one extra step to mark the end of the experiment as shown in Figure 1. The rising slopes of the steps are used to mark the beginning and end of every trial. The EOG data are sliced likewise. A custom event detection algorithm was defined inside the FieldTrip function ft_definetrial to perform this task.

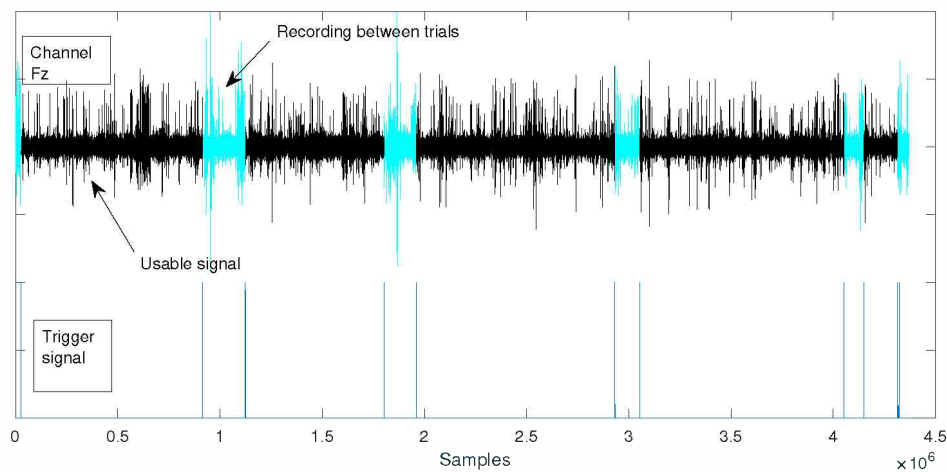

**Figure 1.** Continuously recorded Fz channel along with the auxiliary trigger channel. The onset and offset of every trial are marked with a narrow pulse in the auxiliary trigger channel. An extra pulse marks the end of the experiment session. In this experiment, five different trials were presented to the subject.

### 3.2. Filtering

First of all, note that the EOG and trigger channel were eliminated from the main processing. The raw data was down sampled from 2048 Hz to $f_s = 250$ Hz to lower the data storage and the processing time [9]. The down sampled data were then high-pass filtered with a cut-off frequency $f_{hp} = 1$ Hz, as it is done in [2], where auditory responses measured by conventional and ear-EEG are considered, or in [17], where brain responses to sound are also considered, and low-pass filtered with cut-off frequency $f_{lp} = 47$ Hz (note that for the analysis of higher frequency signals a different filter setting needs to be used). Additionally, a band-stop filter with 49–51 Hz rejection band was applied to strongly reject the 50 Hz power line signal [9,17]. The low-pass and high pass filters are linear phase FIR filters or order $\lfloor 3 * fs/f_{lp/hp} \rfloor$ (where $\lfloor * \rfloor$ means round to the nearest integer towards zero) obtained as a least squares approximation to the desired frequency response windowed using a Hamming window [19,20]. The band-stop filter is an order 4 Butterworth filter; in this case, the presence of zeroes on the unit circle in the z-plane and close poles makes a steep transition. The filters are applied to the EEG signals in a forward and backward direction; this non-causal strategy has the effect of removing the delay caused by the filtering process (if the trailing samples are carefully removed) and doubling the actual filters' order.

### 3.3. Eye Artifact Annotation

Eye artifacts have been manually identified and a membership function (MSF) similar to the one described in [21] was created on the basis of the annotations. The MSF is a logical sequence defined with the same length as the recorded data samples in which the artifacts are marked with the value $+1$ and other samples are marked with 0.

To this end, a tool was built to allow the annotation and later correction of the marks that identify the artifacts. A snapshot of the tool developed is shown in Figure 2. This figure shows the channels considered by the analysts to perform the annotation, which include the available EOG channel and a number of other EEG channels selected on the basis of their location and correlation to the EOG channel.

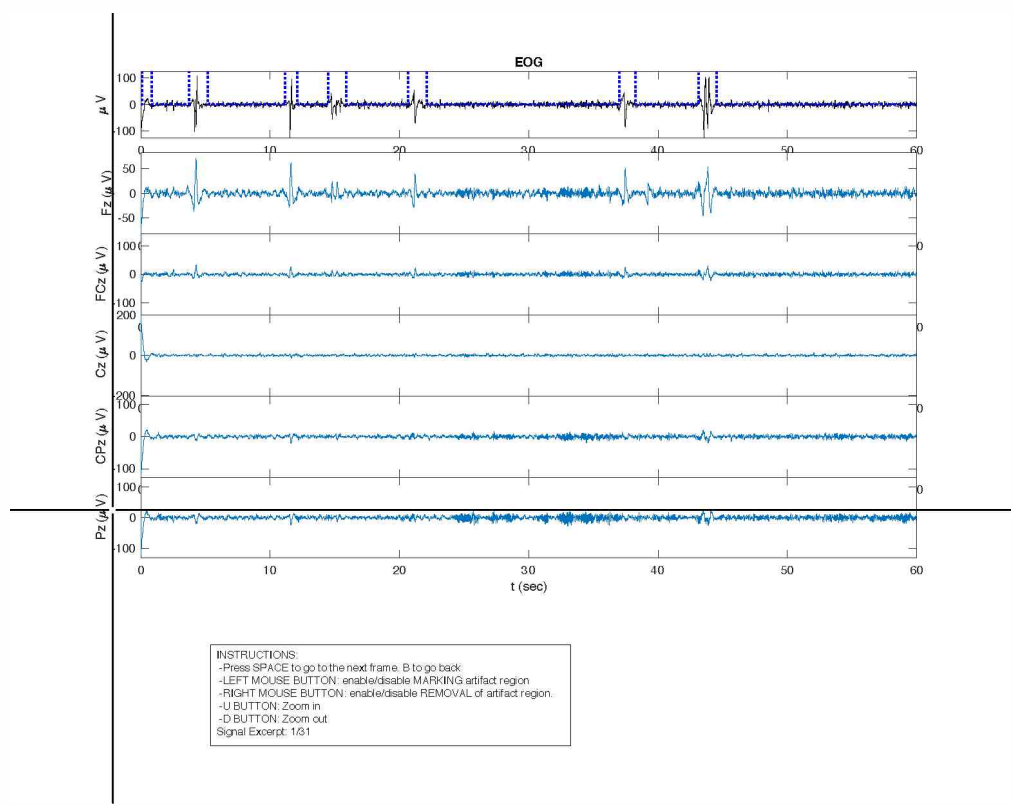

**Figure 2.** Snapshot of the tool developed for manual eye artifact annotation; the cross is used to mark/unmark excerpts of arbitrary duration. Excerpts are marked over the EOG channel plot.

This tool allows one to move between different sections of the recorded signal, mark signal excerpts with eye artifacts, and easily modify the annotation to enlarge, reduce, add, or remove artifactual excerpts by observing at the same time the EOG channel available and close EEG channels.

## 4. Artifact Processing Methods

In this section, the methods proposed to diminish the influence of eye artifacts in the analysis of EEG signals are described. These methods rely on the utilization of ICA for the analysis, so a brief introduction to ICA is done first. Then, the artifact processing methods devised are described.

### 4.1. Independent Component Analysis

Independent component analysis (ICA) is employed to obtain independent components (ICs) excluding the EOG channel [9]. ICA decomposes the initial signals $\mathbf{x}(t) = [x_1(t), x_2(t), \ldots, x_n(t)]^T$ into new statistically independent sources $\mathbf{s}(t) = [s_1(t), s_2(t), \ldots, s_n(t)]^T$ through an unmixing matrix $\mathbf{W}$ such that:

$$\mathbf{s}(t) = \mathbf{W} \cdot \mathbf{x}(t) \tag{1}$$

Then, resulting independent components represent underlying cerebral and non-cerebral sources [13].

### 4.2. Rejection of ICA Components

Following the idea of ICA in the EEG analysis context, we will identify components or artifacts in specific components for their removal or attenuation. The trial subdivision of the signals is considered irrelevant regarding ICA. Nevertheless, the resulting $\mathbf{s}(t)$ maintains the trial substructure so that each IC is defined as $s_n(t) = [s_{n,1}(t), s_{n,2}(t), \ldots, s_{n,m}(t)]$ where $m$ is the number of trials.

In the literature, clean EEG data are obtained by simply removing components, which were previously selected via visual inspection of the data [9,11], automatically by template matching [13] or by EOG cross-correlation measures [14].

In this paper, manual IC identification is not done but the automatic identification of components and their artifactual excerpts is considered by measuring the cross-correlation coefficient with the EOG channel based on the fact that the EOG channel contains primarily ocular motion.

Possible small delays between the recorded signals are considered though these should not be significant [3]. Thus, it is expected to find the most prominent peak around $\tau = 0$; for most cases, the peak was found within $\pm 20$ ms from that point, which, at the considered sampling rate, means a slack of less than 7 samples. The cross-correlation function is normalized by the variances of the signals involved in the calculation. Note that only negative correlation values are valid, since the polarity of eye artifact potentials are inverted in the scalp electrodes [1]. Hence, the minimum cross-correlation coefficient for each component was chosen.

The negative cross-correlations coefficients $\rho$ between the corresponding EOG trial and each trial of each IC are computed as follows:

$$\rho(n, m) = \frac{\min\limits_{|\tau| < 7} \left( \sum\limits_{i=-\inf}^{\inf} EOG_m(i) s_{n,m}(i - \tau) \right)}{\sqrt{\sigma^2_{EOG_m} * \sigma^2_{s_{n,m}}}} \tag{2}$$

This produces a $n \times m$ matrix of negative cross-correlation coefficients $\mathbf{C}$ for each participant:

$$\mathbf{C} = \begin{bmatrix} \rho(1,1) & \rho(1,2) & \ldots & \rho(1,m) \\ \rho(2,1) & \rho(2,2) & \ldots & \rho(2,m) \\ \vdots & \vdots & \ddots & \vdots \\ \rho(n,1) & \rho(n,2) & \ldots & \rho(n,m) \end{bmatrix} \tag{3}$$

where $n$ is the number of different ICs and $m$ is the number of trials.

An illustration of the coefficients of **C** is plotted in Figure 3 where every dashed line corresponds to a different trial for a selected subject.

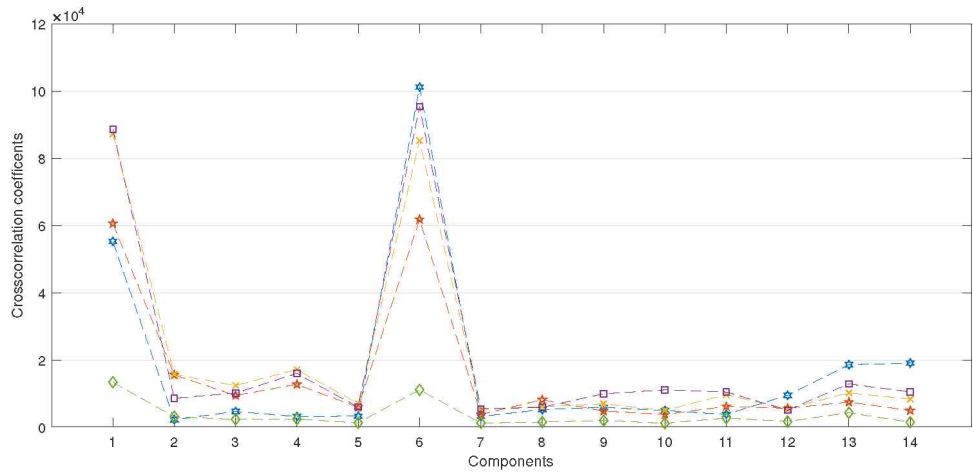

**Figure 3.** Illustration of the cross-correlation coefficients obtained between every IC and EOG in all trials for subject $S_{12}$ (negative values are reversed).

Then, the modulus of the elements of **C** are computed and the rows summed (which represent the trials) according to:

$$\mathbf{cc} = \begin{bmatrix} \sum_{m=1}^{5} |\rho(1,m)| \\ \sum_{m=1}^{5} |\rho(2,m)| \\ \vdots \\ \sum_{m=1}^{5} |\rho(n,m)| \end{bmatrix}^{T} \tag{4}$$

Figure 4 shows an example of the vector **cc** of cumulative cross-correlation coefficients for the same subject as in Figure 3. The two components with the highest values in **cc** are selected as artifactual ones and marked for further processing, i.e., complete removal (CR). This choice was adopted since it was observed that, in the subjects analyzed, two of the components showed a significantly larger cumulative cross-correlation coefficient than the others; see Figure 4 (note that an analysis of this fact with a larger set of subjects would be beneficial for the definition of this or a different approach to select such components).

A scheme of the process described in this section is drawn in Figure 5.

Though processing can be concreted in the complete removal (CR) of selected components, their partial rejection (PR) can also be considered as it will be described in the next section, for the later analysis of reconstructed EEG channels.

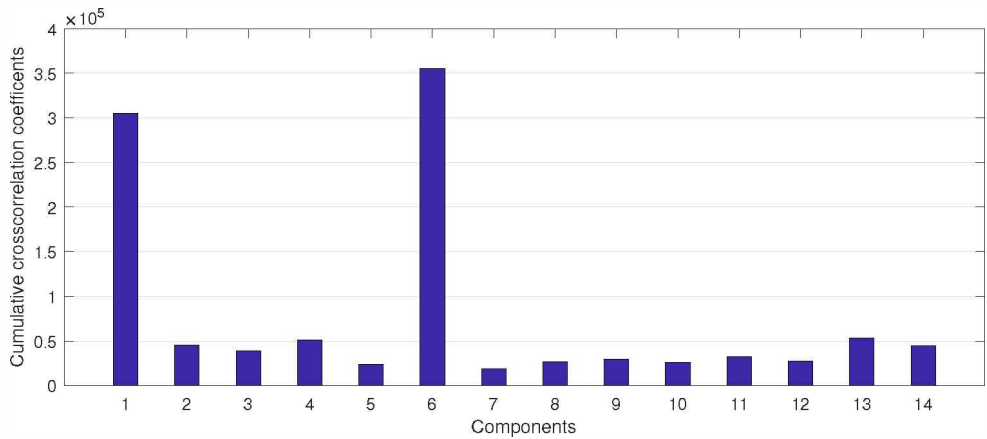

**Figure 4.** Example of cumulative cross-correlation coefficients between every IC and EOG, summed across trials, for subject $S_{12}$.

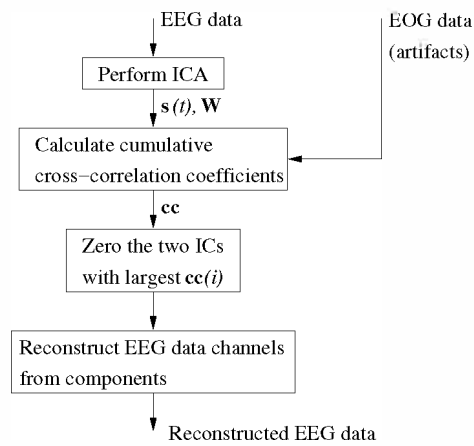

**Figure 5.** Scheme of the complete removal (CR) approach.

## 4.3. Partial Rejection of ICA Components

The partial rejection (PR) of artifactual components is achieved by attenuating the signal excerpts related to eye artifacts but leaving non-artifact samples untouched. The samples related to artifacts are identified by the membership function (MSF) obtained as described in Section 3.3, which is later converted into a Windowed Membership Function (WMSF) by employing smooth transitions adding Blackman's window slopes to the edges of the steps in every marked excerpt. Figure 6 illustrates such a transformation.

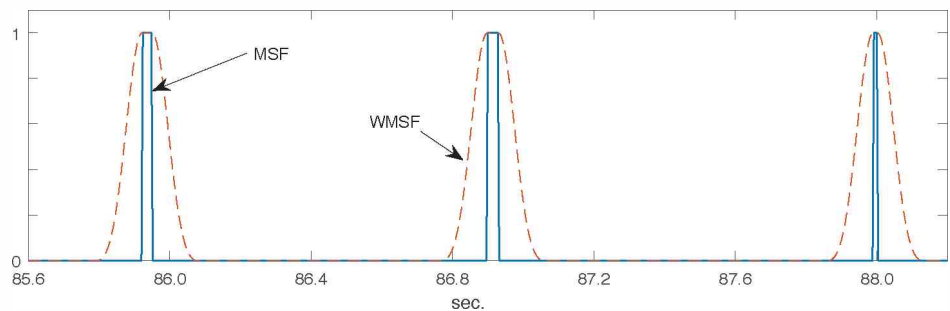

**Figure 6.** A piece of a MSF and its corresponding WMSF.

A MSF can be tuned following this approach by using an attenuation factor, $\alpha$, with a value between 0 and 1 to adjust the desired attenuation strength, with 0 for no attenuation, employing the following expression:

$$s_{\alpha,n}(t) = s_n(t) \cdot (1 - \alpha WMSF_k(t)) \tag{5}$$

In this paper $\alpha$ is fixed to 1 to better analyze the effect of the rejection scheme.

Partial artifact rejection is then performed for each selected component. A graphic representation of the partial rejection produced by the utilization of the WMSF is shown in Figure 7.

A simple scheme of the process described to perform PR of independent components is drawn in Figure 8.

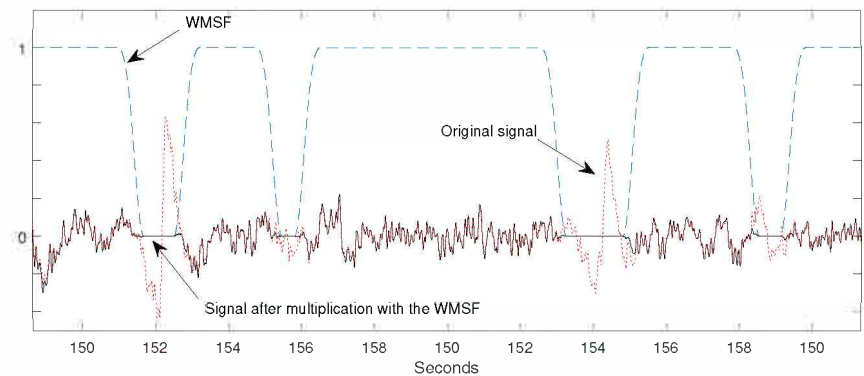

**Figure 7.** Example of the application of a WMSF to a signal component.

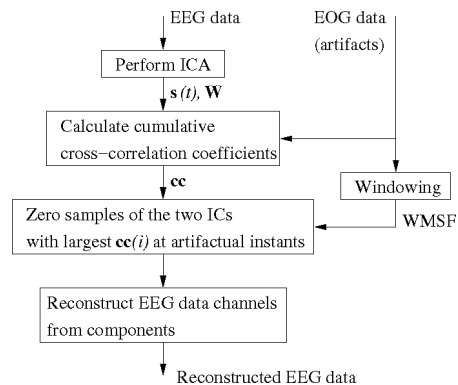

**Figure 8.** Scheme of the partial removal (PR) approach.

### 4.4. Artifact-Diminished ICA Unmixing Matrix

This method to diminish the influence of eye-movement related artifacts performs in a direction different than the previous ones. Now, the extraction of signal components with the influence of artifacts reduced is considered. Thus, the main goal is to generate an unmixing matrix $\mathbf{W}'$ (see Equation (1)) with diminished influence from eye artifacts. In order to achieve this goal, the artifactual excerpts must be removed before searching for the unmixing matrix. Our scheme aimed completely remove the artifactual samples leading to shorter versions of the EEG signal channels. Artifactual samples are identified here by the previously considered, manually annotated MSF.

In summary, ICA was fed with the shorter and artifact-diminished version of the EEG channels, generating an unmixing matrix, $\mathbf{W}'$, more weakly affected by the influence of artifacts, and a new set of independent components $\mathbf{s}'(t)$. This new set of independent components generated should not contain specific components for the eye-artifact sources; obviously, they should not be used for such artifact rejection purposes. However, the independent components obtained by means of this procedure will be well suited to identify other types of artifacts not previously considered [22] or other brain signals; e.g., in [23] or [24] specific brain signals are searched for after ICA.

A simple diagram of the process to obtain the the unmixing matrix $\mathbf{W}'$ is shown in Figure 9.

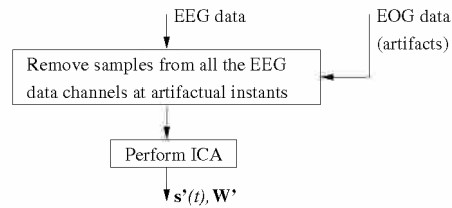

**Figure 9.** Simple diagram of the process to obtain the unmixing matrix $\mathbf{W}'$ with diminished influence of artifacts.

## 5. Results and Discussion

In this section, the results and a discussion on the proposed schemes to diminish the influence of artifacts in later signal analyzes are presented. Note that, although the manuscript is focused on eye artifacts, the ideas on the reduction of the effect of artifacts can be applied to other types of interfering signals.

The number of artifacts manually detected and their main features for each of the subjects are drawn in Table 3. Furthermore, data across all subjects and trials are shown in this table.

**Table 3.** Number of artifacts manually detected and their main features.

| Subject | $S_9$ | $S_{10}$ | $S_{12}$ | $S_{18}$ | $S_{19}$ | Overall |
|---|---|---|---|---|---|---|
| # artifacts | 194 | 260 | 361 | 318 | 253 | 1386 |
| Artifact duration (%) | 27.40 | 46.83 | 50.36 | 62.80 | 30.71 | 43.62 |
| Mean artifact duration (s) | 2.55 | 3.26 | 2.52 | 3.57 | 2.19 | 2.84 |
| Std of artifact duration (s) | 2.57 | 2.98 | 2.22 | 3.44 | 1.71 | 2.72 |
| Median artifact duration (s) | 1.68 | 1.96 | 1.75 | 2.26 | 1.55 | 1.78 |

Note that the relative artifact duration with respect to the total duration of the measurements ('Artifact duration (%)' in Table 3) is high, which indicates that the annotation of artifacts was likely biased to avoid missing artifacts and the marks where not tightly adjusted to the artifacts. This fact is observed to happen coherently across all the subjects; nevertheless, it must be noted the large relative accumulated artifact duration found in subjects $S_{10}$, $S_{12}$, and, especially, $S_{18}$, as well as the large number of artifacts identified, particularly in the cases of subjects $S_{12}$ and $S_{18}$.

### 5.1. Partial Rejection of ICA Components

The method explained in Section 4.3 was applied to the available data. The base method described in Section 4.2 is considered. The cumulative cross-correlation coefficients of each channel with the EOG were computed using Equations (2)–(4) before ICA and again after the reconstruction of the data from the processed ICs.

The negative cross-correlation coefficient with the EOG channel (EOG-correlation for short) is the chosen parameter to obtain a measure of the influence of the EOG signal in other channels, i.e., the effect of eye-related artifacts on them, since the EOG signal primarily contains eye movement energy [14].

Nonetheless, a zero EOG-correlation is neither desired nor actually possible considering that the EOG channel not only contains energy derived from eye movements but also brain activity. Taking into account this observation, the reduction of EOG-correlation can be used as a comparative measure on how good an artifact-cleaning method is against the others, though it never should be used as an absolute measure, since the maximum reduction value is not stated and certainly could not be 100%.

Figure 10 contains an example of how the component removal processes considered affect channel Fz in one of the test subjects.

Table 4 contains the average cumulative cross-correlation coefficients for the EEG channels before ICA and after their reconstruction using complete removal (CR) of components selected by using cross-correlation coefficients with the EOG channel, and the partial IC removal approach (PR) described in Section 4.3, with the MSF described in Section 3.3. The last row summarizes the resulting reduction of the cross-correlation coefficient.

The complete component removal approach produces a large reduction of the cross-correlation coefficients, as expected, but at the same time it removes part of the brain activity as seen in Figure 10, where it is clearly visible that non-artifactual parts are significantly attenuated. This implies that ICA cannot completely separate brain signals from ocular activity or other interfering signals or noise. This is an important observation of an issue that can be due to different causes that might be related to ICA limitations derived from the signal model, measurement constraints (i.e., low-density electrode array), or computation schemes [22]. Thus, a complete removal of components is not considered an optimal choice since ICA cannot actually guarantee that certain components contain only noisy data and do not contain useful pieces of information [9].

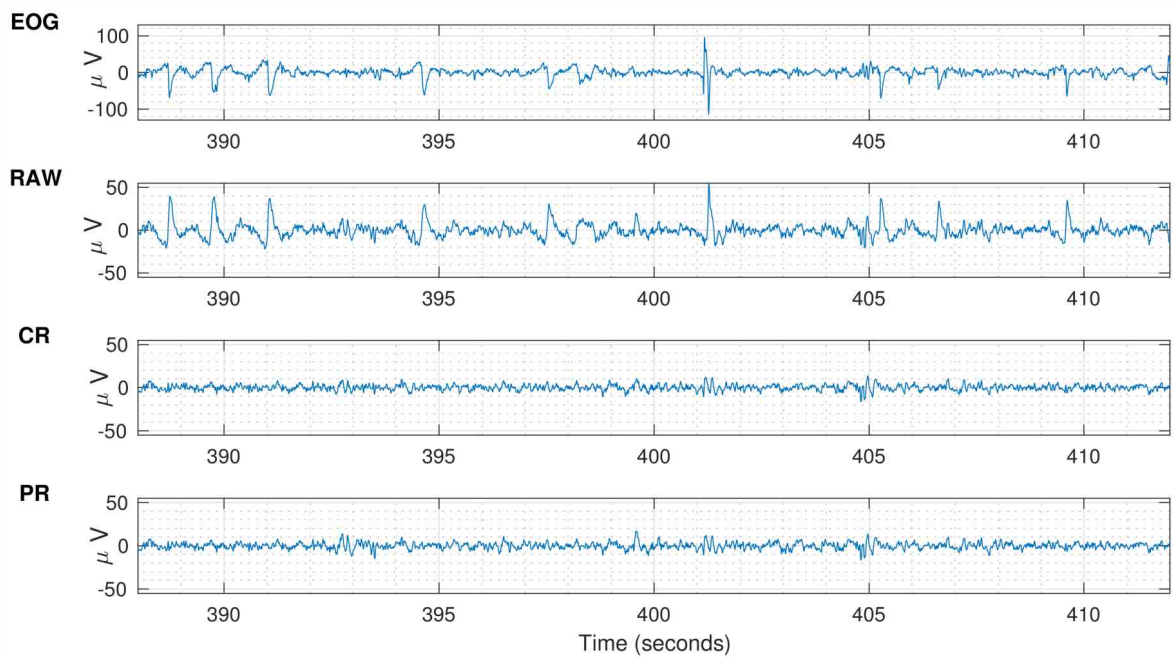

**Figure 10.** Example of signal of channel Fz before and after applying the considered schemes to diminish the influence of artifacts based on ICA analysis and complete IC removal (CR) and partial IC removal (PR) helped by manual artifact annotations.

Regarding the utilization of the proposed method of partial rejection (PR) of ICs, the reduction of the cross-correlation coefficients is also large, comparable to that of the CR method, while better preserving the shape of the signals in non-artifactual excerpts (see Figure 10) at the cost of the necessity of artifact annotation.

We also investigated the effect of complete/partial removal of IC components regarding the induced changes of signal-to-noise ratio (SNR) as defined in [25]. However, note that the stimuli considered greatly differs from the commonly used ones regarding their duration and repetitiveness, implying that a single trial per user for five different experiments can be considered. With this fact in mind, we have followed the procedure in [25] aware of the low SNR values that should be found due to the specificity of the stimuli and signals involved. Nevertheless, it was considered valuable to observe the behavior of the SNR under the proposed scenarios.

**Table 4.** Average cumulative cross-correlation coefficients before and after applying the considered artifact processing schemes based on complete/partial IC rejection. Note that all absolute values in this table are divided by $10^5$.

| Channel | Before ICA | After IC Rejection | |
|---|---|---|---|
| | | Complete (CR) | Partial (PR) |
| 'Fz' | 2.7657 | 0.3279 | 0.3354 |
| 'FCz' | 2.4154 | 0.2192 | 0.2209 |
| 'Cz' | 1.1386 | 0.4025 | 0.5064 |
| 'CPz' | 0.9353 | 0.3702 | 0.4553 |
| 'Pz' | 0.9562 | 0.2161 | 0.2582 |
| 'POz' | 0.8500 | 0.3119 | 0.3027 |
| 'F3' | 2.7674 | 0.3168 | 0.2911 |
| 'F4' | 1.9507 | 0.3080 | 0.2769 |
| 'C3' | 1.4484 | 0.3923 | 0.4330 |
| 'C4' | 1.1308 | 0.3498 | 0.3683 |
| 'T7' | 0.5083 | 0.2243 | 0.2378 |
| 'T8' | 0.8184 | 0.4198 | 0.4529 |
| 'P3' | 0.8772 | 0.2856 | 0.3490 |
| 'P4' | 0.8021 | 0.2833 | 0.3026 |
| Reduction | | 77.13% | 72.68% |

Table 5 shows that the SNR values obtained were low, as expected, however, they grew with the utilization of the contemplated schemes, achieving their largest increase with the utilization of the CR scheme, though the PR method performed close. The complete rejection scheme achieved a global increase of 16.27% of SNR, whereas the PR method attained an increase of 12.37%.

**Table 5.** SNR before and after CR/PR component removal to diminish the influence of artifacts. Global experiment makes reference to the consideration of all the experiments as a whole for the calculation of SNR.

| Experiment | Before ICA | After IC Rejection | |
|---|---|---|---|
| | | Complete (CR) | Partial (PR) |
| The Magic Flute | 0.3328 | 0.3860 | 0.3754 |
| Emperor Waltz | 0.3361 | 0.3986 | 0.3873 |
| Universal Mind | 0.3522 | 0.4108 | 0.3846 |
| Adios Nonino | 0.3474 | 0.3983 | 0.3919 |
| Speech | 0.3423 | 0.3983 | 0.3905 |
| Global | 0.3429 | 0.3987 | 0.3853 |

### 5.2. Artifact-Diminished ICA Unmixing Matrix

In this case, we specifically altered the result of the ICA by modifying the actual input data. Thus, special attention will be paid to the influence of the proposed method on the unmixing matrix and observe the changes in it.

Figure 11a shows the original unmixing matrix, $\mathbf{W}$, and Figure 11b the artifact-diminished unmixing matrix, $\mathbf{W}'$, obtained as described in Section 4.4, with eye-artifacts manually detected. Figure 12a contains the difference matrix $\mathbf{D}$ between the matrices in Figure 11, where every element has been calculated as follows:

$$\mathbf{D}(n, m) = \mathbf{W}'(n, m) - \mathbf{W}(n, m) \qquad (6)$$

Figure 12b displays the logarithmic relative difference matrix $\mathbf{D}_{L_R}$ obtained using:

$$\mathbf{D}_{L_R}(n,m) = \log_{10}\left(\left|\frac{\mathbf{D}(n,m)}{\mathbf{W}(n,m)}\right|\right) \tag{7}$$

Note that, in order to obtain a meaningful difference matrix, the vectors that perform the transformation in the unmixing matrices $\mathbf{W}$ and $\mathbf{W}'$ in Figure 11 are coherently sorted in descending order of their norm. In the illustrations, component 1 refers to the vector with the largest norm, whereas component 14 corresponds to the one with the smallest norm.

It can be observed that the unmixing transformation changed substantially when ICA was fed with artifact-diminished data instead of the original data. Thus, it is consequently expected that the components obtained by the devised transformation scheme will represent the underlying brain sources more accurately.

Moreover, the images in Figure 12 clearly draw the changes between $\mathbf{W}$ and $\mathbf{W}'$ which can be positive when a certain contribution of a certain channel in certain component rises and negative otherwise, see Figure 12a,b renders the relative magnitude of the differences. A value of 0 corresponds to no variation of the magnitude. However, it is observed that many elements suffer a change of their magnitude over the 100% of their magnitude. This observation was noticed across all the subjects and is a remarkable illustration of the importance of the preprocessing of artifacts before EEG analyzes.

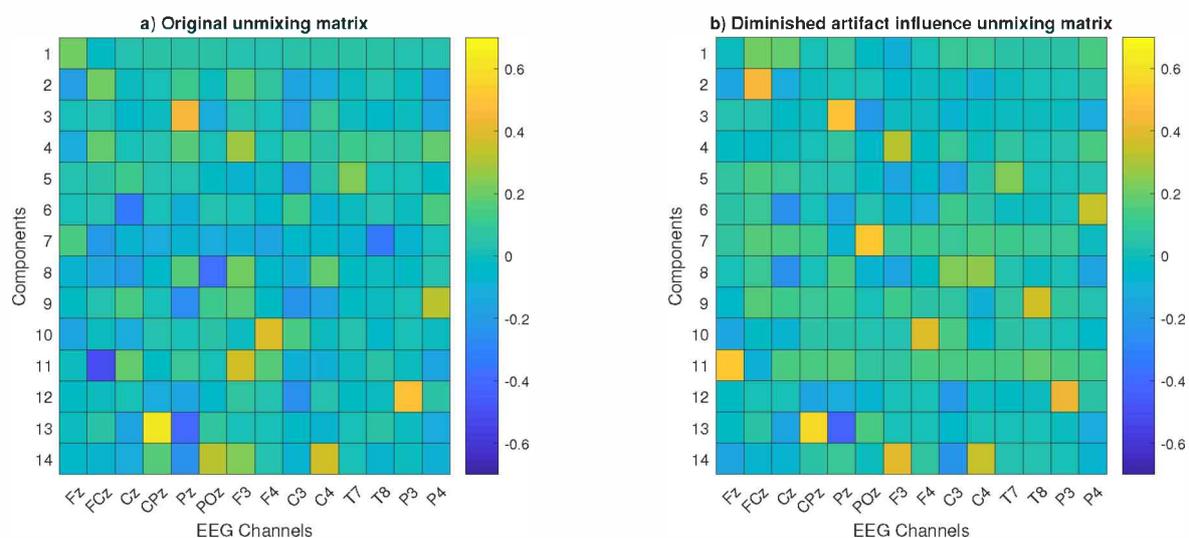

**Figure 11.** (**a**) Unmixing matrix for ICA decomposition. (**b**) Unmixing matrix for ICA decomposition obtained from data with the influence of eye-artifacts diminished, for subject $S_{19}$, according to the procedure in Section 4.4.

Figure 13 illustrates the topography of the independent components corresponding to the conventional unmixing matrix with no eye-artifact processing and the artifact-diminished unmixing matrix corresponding to the handmade eye-artifact detections and processing scheme for subject $S_{19}$. Clear differences can be observed in the distribution of the components' topography in this example. The component numbers are unrelated across the two images.

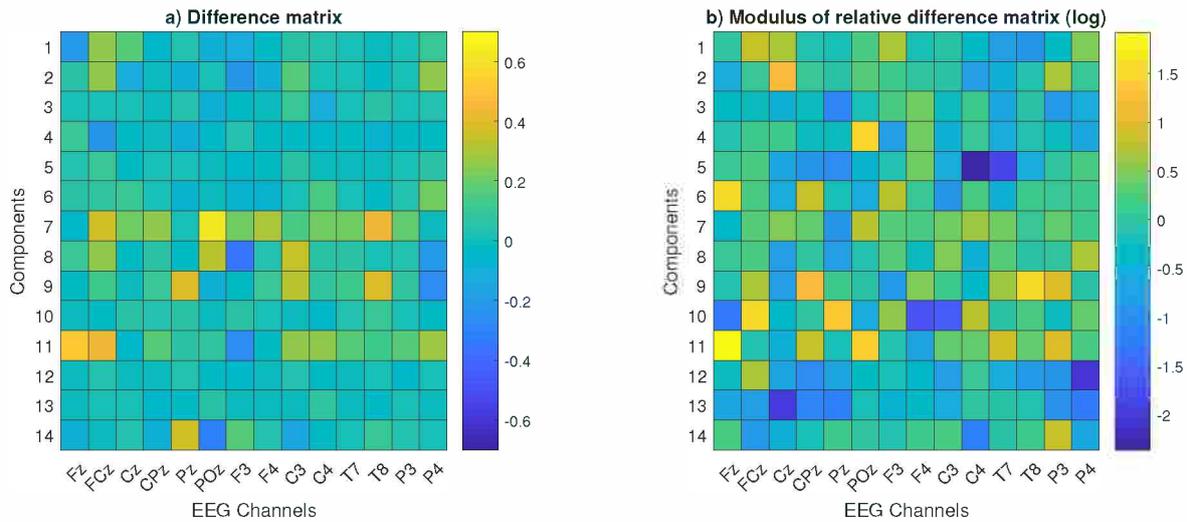

**Figure 12.** (**a**) Difference matrix between the original unmixing matrix given by ICA decomposition and the artifact-diminished unmixing matrix shown in Figure 11. (**b**) Relative difference matrix in logarithmic scale.

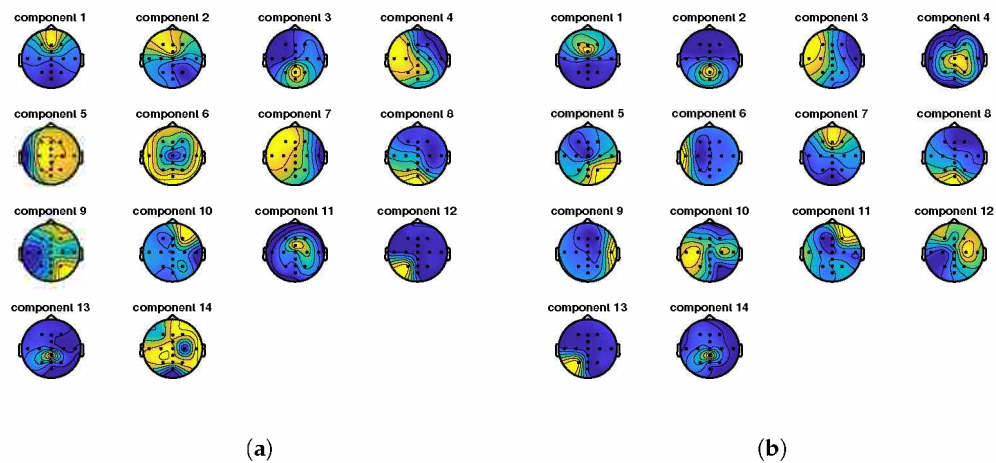

(**a**) (**b**)

**Figure 13.** Illustration of spatial topography differences of the components obtained by applying the unmixing matrices **W** (**a**) and **W**′, with handmade detection of eye-artifacts (**b**).

## 6. Conclusions

This paper presented a study on the importance of specific artifact preprocessing schemes for EEG analyzes when trials cannot be simply discarded.

Novel schemes were presented to diminish the influence of eye movement and blinks for the analysis of EEG data by proposing partial rejection of independent components based on artifact detection, in contrast to the widely used complete removal approach. Complete removal of components is not considered an optimal choice in the chosen context since ICA cannot guarantee that certain components contain only noisy data and do not contain useful pieces of information [9].

Partial rejection of components relies on the utilization of an artifact detection procedure to eliminate eye-related artifacts and, at the same time, minimize the loss of valid data in the EEG. In order to assess the performance of the partial rejection approach, the reduction of EOG-correlation was computed and the results were compared against those corresponding to the complete removal of independent components.

Although the reduction rate produced by complete removal of ICs is higher, close reduction rates can be achieved by the proposed partial rejection scheme using manual eye-artifact annotations, as performed in this work.

Finally, this paper introduced a new different method to cope with the influence of artifacts in ICA decomposition. Specifically, feeding the ICA algorithm with a shorter, artifact-diminished version of the EEG channels aimed at obtaining an unmixing matrix and ICs that benefit from reduced influence of eye artifacts. An analysis of its importance in the definition of the signal transformation was also provided. It is remarkable that the fact that the difference between the original and artifact-diminished unmixing matrices is significant, which states the importance of rejecting the artifactual excerpts in the data. Nevertheless, constraints and measurement conditions must be taken into account, specifically, the low number of electrodes available, which constitutes a relevant constraint for the signal model and the analysis performed, and the number of subjects; the increase of the number of subjects would benefit the extraction of conclusions and their general applicability.

**Author Contributions:** Conceptualization: A.V., L.J.T., I.B., A.M.B., E.B. and N.T.H.; Data curation: A.V. and N.T.H.; Formal analysis: A.V., L.J.T., I.B., A.M.B., E.B. and N.T.H.; Funding acquisition: L.J.T., I.B. and E.B.; Investigation: A.V., L.J.T., I.B., A.M.B., E.B. and N.T.H.; Methodology: A.V., L.J.T., I.B., A.M.B. and E.B.; Software: A.V., L.J.T., I.B., A.M.B. and N.T.H.; Supervision: L.J.T., I.B. and E.B.; Validation: E.B. and N.T.H.; Visualization: A.V., L.J.T., A.M.B. and N.T.H.; Writing original draft: A.V., L.J.T., A.M.B. and I.B.; Writing review and editing: A.V., L.J.T., I.B., A.M.B., E.B. and N.T.H.

**Funding:** This work has been funded by the Ministerio de Economía y Competitividad of the Spanish Government under Project No. TIN2016-75866-C3-2-R. This work has been done at Universidad de Málaga, Campus de Excelencia Internacional (CEI) Andalucía TECH. Center for Music in the Brain is funded by the Danish National Research Foundation (DNRF117).

**Conflicts of Interest:** The authors declare no conflict of interest. The founding sponsors had no role in the design of the study; in the collection, analyzes, or interpretation of data; in the writing of the manuscript, and in the decision to publish the results.

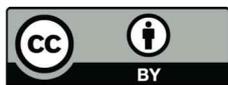